\documentclass[journal]{IEEEtran}

\usepackage{amsmath,amssymb}
\usepackage{colortbl}
\usepackage{diagbox}
\usepackage{subeqnarray}
\usepackage{cases}
\usepackage{setspace}
\usepackage{tabularx}
\usepackage{booktabs}
\usepackage{multicol}
\usepackage{dcolumn}
\usepackage{graphicx}
\usepackage{subfigure}
\usepackage{float}
\usepackage{epsfig}
\usepackage{amsmath}
\usepackage{amssymb}
\usepackage{array}
\usepackage{epstopdf}
\usepackage{color}
\usepackage{xcolor}
\usepackage[nocompress]{cite}
\usepackage{psfrag}
\usepackage{cite}
\usepackage{multirow}
\usepackage{color}
\usepackage{bm}
\usepackage{adjustbox}
\usepackage{overpic}
\usepackage{rotating}

\graphicspath{{figures/}}
\usepackage[bookmarks,colorlinks,linkcolor=red, anchorcolor=blue, citecolor=green]{hyperref}
\begin{document}

\title{Flexible Image Denoising with \\Multi-layer Conditional Feature Modulation }

\author{Jiazhi~Du,~
        Xin~Qiao,~
        Zifei~Yan,
        Hongzhi~Zhang,
        and~Wangmeng~Zuo,~\IEEEmembership{Senior~Member,~IEEE}
\thanks{This project is partially supported by the National Natural Scientific
Foundation of China (NSFC) under Grant No.s 61671182 and U19A2073.}
\thanks{J. Du, X. Qiao, Z. Yan and W. Zuo are with the Harbin Institute of Technology, Harbin, 150001, China. e-mail: dujiazhi888@foxmail.com; qiaoxin182@gmail.com; amber\_1980@163.com; zhanghz0451@gmail.com; wmzuo@hit.edu.cn.}
}%

\markboth{Journal of \LaTeX\ Class Files,~Vol.~xx, No.~x, xxx~20xx}%
{Shell \MakeLowercase{\textit{et al.}}: Bare Demo of IEEEtran.cls for IEEE Journals}
\maketitle

\begin{abstract}
For flexible non-blind image denoising, existing deep networks usually take both noisy image and noise level map as the input to handle various noise levels with a single model.
However, in this kind of solution, the noise variance (i.e., noise level) is only deployed to modulate the first layer of convolution feature with channel-wise shifting, which is limited in balancing noise removal and detail preservation.
In this paper, we present a novel flexible image denoising network (CFMNet) by equipping an U-Net backbone with multi-layer conditional feature modulation (CFM) modules.
In comparison to channel-wise shifting only in the first layer, CFMNet can make better use of noise level information by deploying multiple layers of CFM.
Moreover, each CFM module takes convolutional features from both noisy image and noise level map as input for better trade-off between noise removal and detail preservation.
Experimental results show that our CFMNet is effective in exploiting noise level information for flexible non-blind denoising, and performs favorably against the existing deep image denoising methods in terms of both quantitative metrics and visual quality.

\end{abstract}


\IEEEpeerreviewmaketitle

\section{Introduction}

\IEEEPARstart{I}{mage} denoising, with the aim of recovering the latent clean image from its noisy observation, is a classical yet active topic in low level vision~\cite{chen2018image,santhanam2016generalized}.
Driven by the progress in network architecture and learning algorithm, convolutional neural networks (CNNs)~\cite{zhang2017beyond} have received much recent research interest in image denoising.
Nowadays, several representative deep denoising networks, e.g., DnCNN~\cite{zhang2017beyond}, FFDNet~\cite{zhang2018ffdnet}, MemNet~\cite{tai2017memnet} and MWCNN~\cite{liu2018multi}, have been suggested and achieved superior performance against traditional model-based methods such as BM3D~\cite{Kostadin2007Image} and WNNM~\cite{gu2014weighted}.

Considering that various real-world noisy images may correspond to different noise levels, flexible image denoising is also a critical issue for practical applications.
Most deep denoising networks, however, still lack flexibility in dealing with additive white Gaussian noise (AWGN) with various noise levels and even spatially variant noise.
For example, non-blind DnCNN~\cite{zhang2017beyond} generally learns a specific model for each noise level.
For handling AWGN with a large range of noise levels (e.g., $[0, 50]$), a large number (e.g., $25$ in~\cite{zhang2017beyond}) of denoising models are usually required to be trained and stored in advance, thereby limiting their practicality in versatile denoising tasks.
Benefited from the powerful modeling ability of CNNs, it is also feasible to train a single deep blind denoising model (e.g., DnCNN-B~\cite{zhang2017beyond}) for dealing with AWGN with a range of noise levels~\cite{Mao2016Image,tai2017memnet}.
However, the learned model is very limited in handling real-world non-Gaussian noise, and even generalizes poorly to AWGN out of the preset noise level range.

Recently, several CNN denoisers have been suggested to address the flexibility issue of non-blind image denoising.
Zhang et al.~\cite{zhang2018ffdnet} present a fast and flexible denoising convolutional
network (i.e., FFDNet).
Given a noisy image with specific noise level, FFDNet simply takes both noisy image and noise level map as the input, and thus can remove AWGN with different noise levels by only using a single network.
The input noise level also plays the role of controlling the tradeoff between noise reduction and detail preservation, making FFDNet practically well on some real-world noisy images~\cite{zhang2018ffdnet}.
For burst denoising, Mildenhall et al.~\cite{mildenhall2018burst} show that taking noise level as input is beneficial to the noise not in the preset noise level range.
Moreover, CBDNet~\cite{guo2019toward} incorporates noise level estimation and non-blind denoising for handling real-world noisy photography, where the later takes noise level map and noisy image as input for the generalization beyond the noise model.
These methods, however, simply concatenate the noise level map and noisy image in the input layer, which are insufficient in modeling their sophisticated interplay and perform limited in balancing noise removal and detail preservation.

\begin{figure*}[htbp]
\centering
\hspace{-0.4cm}
\subfigure[Noisy image ($\sigma_{\text{gt}}=60$)]{
\includegraphics[width=0.245\textwidth]{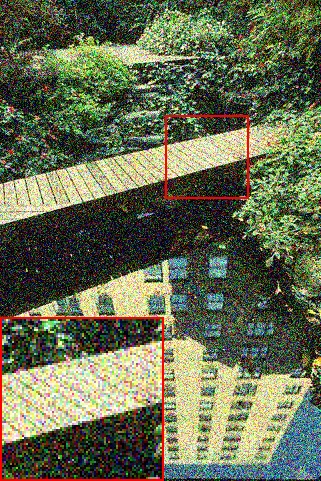}
}\hspace{-2.6mm}
\subfigure[FFDNet ($\sigma_{\text{in}}=55$)]{
\includegraphics[width=0.245\textwidth]{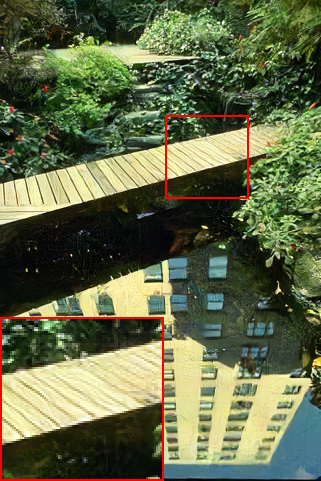}
}\hspace{-2.6mm}
\subfigure[FFDNet ($\sigma_{\text{in}}=60$)]{
\includegraphics[width=0.245\textwidth]{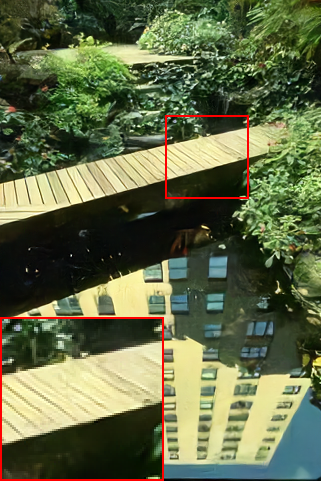}
}\hspace{-2.6mm}
\subfigure[CFMNet ($\sigma_{\text{in}}=60$)]{
\includegraphics[width=0.245\textwidth]{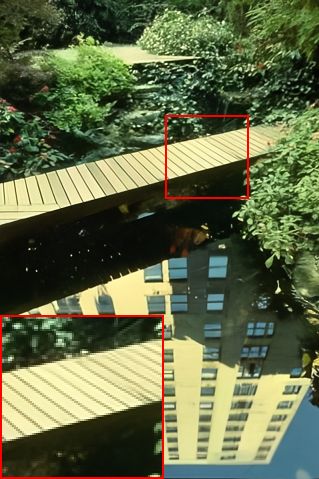}
}\hspace{-2.6mm}

\caption{Denoising results of image \emph{148026} from the CBSD68 dataset with noise standard deviation 60 by FFDNet and our CFMNet. FFDNet ($\sigma_{\text{in}}=60$) is effective in removing noise but may smooth out small-scale details.
FFDNet ($\sigma_{\text{in}}=55$) can preserve more details but may retain some noise in the result.
In comparison, our CFMNet ($\sigma_{\text{in}}=60$) achieves better trade-off between noise removal and detail preservation.}
\label{example}
\end{figure*}

Fig.~\ref{example}(a) shows a noisy image by adding AWGN with noise standard deviation $60$.
Fig.~\ref{example}(b)(c) provide the denoising results by FFDNet with the input noise levels 55 and 60.
It can be seen that FFDNet with the input noise level 60 is effective in removing noise, but may smooth out some small-scale details (see Fig.~\ref{example}(c)).
In comparison, FFDNet with the input noise level 55, i.e., FFDNet ($\sigma_{\text{in}}=55$), can preserve more details but some noise is still retained in the result (see Fig.~\ref{example}(b)).
To sum up, FFDNet is still not sufficient in exploiting noise level information to distinguish between signal and noise, leaving some leeway for better modeling the interplay between noise level map and noisy image.

In this paper, we present a novel flexible image denoising network (i.e., CFMNet) for properly incorporating noise level information into the denoising process.
To begin with, we revisit the input concatenation of noise level map and noisy observation~\cite{zhang2018ffdnet,guo2019toward} from the perspective of feature modulation, and regard them as a channel-wise shifting modulation conditioned by noise level in the first convolution layer.
From this perspective, the limitation of input concatenation can be ascribed to the simple and single layer feature modulation.

To overcome the limitation of input concatenation, we present a novel flexible non-blind image denoising network by considering two aspects.
First, an U-Net backbone is equipped with multi-layer conditional feature modulation (CFM) modules, resulting in our CFMNet.
Instead of channel-wise shifting modulation only in the first layer as FFDNet does, we adopt the residual shifting-based conditional feature modulation (RS-CFM) and deploy it on multiple layers.
In particular, RS-CFM can be treated as a kind of conditional modulation where external conditional information is introduced to modulate convolution activations.
%
%
Existing conditional modulation methods involves conditional normalization~\cite{ioffe2015batch,ulyanov2016instance} and normalization-free modulation~\cite{Perez2017FiLM,wang2018recovering}, and have been widely applied to style transfer~\cite{jing2018stroke}, image-to-image translation~\cite{isola2017image}, visual question answering (VQA)~\cite{de2017modulating}, and single image super-resolution (SISR)~\cite{wang2018recovering}.
Instead of affine transformation adopted in most conditional modulation methods, we find that our residual shifting-based CFM works well for flexible image denoising.
As opposed to input concatenation in~\cite{zhang2018ffdnet,mildenhall2018burst,guo2019toward}, we deploy multiple layers of RS-CFM for better exploiting noise level information to enhance denoising performance.

Furthermore, for better tradeoff between noise removal and detail preservation, our RS-CFM takes convolutional activations from both noisy image and noise level map as input to generate spatially variant shifting map.
From Fig.~\ref{example}(b), it can be seen that FFDNet with smaller noise level is effective in detail preserving but may retain some noise in the salient and smooth regions.
While FFDNet with higher noise level may suffer from removing noise at the potential cost of smoothing out small-scale details (Fig.~\ref{example}(c)).
Thus, one plausible solution is to increase the input noise level for smooth regions and decrease it for the regions with detailed textures.
To this end, we take the convolution activations from both noise level map and noisy image as the input of each RS-CFM, and learn to generate proper spatially variant shifting map for boosting denoising performance.
The convolution activations from noisy image provide the information of image content which can collaborate with noise level map to achieve better tradeoff between noise removal and detail preservation.

Extensive experiments are conducted to evaluate the proposed CFMNet for Gaussian denoising.
The results show that multi-layer RS-CFM is beneficial to denoising performance quantitatively and qualitatively.
And more small-scale details can be preserved by taking convolutional activations from noisy image and noise level map as input to CFM.
In terms of both quantitative metrics and visual quality, our CFMNet performs favorably against the state-of-the-art traditional methods (e.g., BM3D) and deep image denoising methods (e.g., DnCNN~\cite{zhang2017beyond}, MemNet~\cite{tai2017memnet}, FFDNet~\cite{zhang2018ffdnet}, NLRN~\cite{liu2018non} and MWCNN~\cite{liu2018multi}).
To sum up, the merits of performance and flexibility make our CFMNet very competitive in image denoising.

The contributions of this work are summarized as follows:
\begin{itemize}
  \item A novel flexible non-blind image denoising network, i.e., CFMNet, is presented by incorporating multi-layer CFM modules and the U-Net backbone.
      In comparison to input concatenation~\cite{zhang2018ffdnet}, our CFM adopts residual shifting-based modulation and is deployed on multiple layers to enhance denoising performance.
  \item In each CFM, the convolutional activations from both noisy image and noise level map are taken as input to generate proper shifting map, and better tradeoff between noise removal and detail preservation can be achieved by our CFMNet.
  \item Experimental results demonstrate the denoising performance and flexibility of our CFMNet.
      In terms of quantitative metrics and visual quality, our CFMNet performs favorably against the state-of-the-art methods, making it highly attractive in image denoising.
\end{itemize}

The remainder of the paper is organized as follows.
Section~\ref{sec:related_work} gives a brief survey on deep image denoising and conditional feature modulation.
Section~\ref{sec:method} presents our CFMNet for flexible image denoising.
Subsequently, Section~\ref{sec:experiments} reports the experimental results and Section~\ref{sec:conclusion} concludes this work with several concluding remarks.

\section{Related Work}
In this section, a brief survey is first given to the recent progress in deep denoising networks.
Naturally, flexible non-blind denoising can be treated as the problem of plain image denoising conditionally modulated by noise level map.
Thus, we also briefly review the relevant feature modulation methods with conditional information.
\label{sec:related_work}
\subsection{Deep Image Denoising Networks}
Recent years have witnessed the unprecedented success of deep CNNs in image denoising.
Early deep image denoising models usually adopt stacked denoising auto-encoders~\cite{vincent2008extracting} and CNNs~\cite{jain2009natural}, but cannot compete with the highly effective traditional methods such as BM3D~\cite{Kostadin2007Image}.
Burger et al.~\cite{burger2012image} adopted plain multi-layer perception (MLP) to learn the denoising mapping on image patches, and achieved promising performance in comparison to BM3D.
Recently, by incorporating residual learning and batch normalization (BN), Zhang et al.~\cite{zhang2017beyond} developed a deep denoising network DnCNN which outperforms the benchmark BM3D with a large margin.
Mao et al.~\cite{Mao2016Image} suggested a deep RED-Net by adding skip connections symmetrically to a fully convolutional network.
Subsequently, the performance of deep denoising networks has been continuously improved with the introduction of recursive units~\cite{tai2017memnet}, multi-scale memory system~\cite{jia2019focnet}, U-Net~\cite{Mao2016Image}, 2D Wavelet transform~\cite{liu2018multi}, feature attention~\cite{anwar2019real}, densely connected hierarchical learning~\cite{park2019densely} and top-down self-guidance~\cite{gu2019self}.

The non-local self-similarity within an image has been widely used in most promising traditional model-based denoising methods~\cite{buades2008nonlocal,Kostadin2007Image,gu2014weighted}, and can also be leveraged to collaborate with neural network.
Yang and Sun~\cite{yang2017bm3d} suggested to learn domain transform and activation functions for BM3D.
And non-local module is also used as a pre-processing step in several cascaded learning models~\cite{Qiao2017,Lefkimmiatis2017}.
However, the non-local matching in these methods is hand-crafted and non-differential, and is conducted in fixed feature space.
Wang et al.~\cite{wang2018non} suggested an end-to-end trainable non-local neural network for image and video classification.
As for image denoising, Liu et al.~\cite{liu2018non} integrated non-local module into recurrent networks (NLRN) to model the feature correlation between each location and its limited neighborhood.
Going beyond weighted averaging, Pl\"{o}tz and Roth~\cite{Pl2018Neural} presented the neural nearest neighbors block to relax $k$-nearest neighbors (KNN) selection for image denoising.

While achieving promising performance, the above mentioned methods generally suffer from inflexibility in handling various noise levels or spatially variant noise.
One possible solution is to learn a blind denoising model for a range of noise levels~\cite{zhang2017beyond,Mao2016Image,tai2017memnet}, but the learned model usually suffers from the lack of generalization ability to real-world noise and even AWGN not in the preset range.
Recently, it has shown that the flexibility and generalization issue can be well addressed by input concatenation, i.e., taking both noisy image and noise level map as the network input \cite{zhang2018ffdnet,mildenhall2018burst,guo2019toward}.
However, input concatenation can be treated as a specific shifting-based modulation in the first convolution layer, which is still not sufficient in removing noise removal while preserving fine details.
Instead of input concatenation, this paper presents a multi-layer RS-CFM modulation modules for capturing sophisticated interplay between noisy image and noise level, and leverages spatially variant shifting map to balance noise removal and detail preservation.

\subsection{Feature Modulation with Conditional Information}
{The feature maps in the hidden layers of CNN can be regarded as a kind of generic representation of input image.
Feature modulation can then been exploited to either boost CNN training or tailor the model to specific conditional information.}

Conventional feature normalization modules, e.g., batch normalization (BN)~\cite{ioffe2015batch} and instance normalization (IN)~\cite{ulyanov2016instance}, generally do not rely on conditional information and have been extensively adopted in modern deep networks.
Nonetheless, for several vision tasks such as image-to-image translation~\cite{isola2017image} and visual question answering~\cite{de2017modulating}, external conditional information can be introduced to generate desired or better solutions.
For incorporating conditional information, feature normalization has been extended to conditional normalization, e.g., conditional BN~\cite{de2017modulating} and adaptive instance normalization (AdaIN)~\cite{huang2017arbitrary}.
In each conditional normalization layer, convolutional activations are first normalized to normal distribution, and conditional information is then leveraged to learn affine transformation (scaling and shifting) for modulating activations.
Besides, normalization-free modulations, e.g., feature-wise linear modulation (FiLM)~\cite{Perez2017FiLM} and spatial feature transform (SFT)~\cite{wang2018recovering}, are also suggested for visual question answering and semantic map guided SISR.
Spatially invariant affine transform usually is adopted in most conditional normalization and normalization-free modulation modules.
While SFT~\cite{wang2018recovering} and spatially-adaptive normalization (SPADE)~\cite{park2019semantic} suggest to learn spatially variant affine transform for enhancing visual quality.

In this work, we also adopt spatially variant transform for conditional feature modulation.
Instead of affine transformation involving both scaling and shifting, we find that our residual shifting empirically works well for our task.
In contrast to existing methods, we consider convolutional activations from both noise image and noise level map to generate spatially variant transform for better tradeoff between noise removal and detail preservation.

\section{Proposed Method}
%
%
Suppose $\bm{y}$ be a noisy image with AWGN, $\sigma^2$ be the noise variance, and $\bm{x}$ be the corresponding clean image.
Flexible non-blind image denoising is then defined as the restoration of clean image $\bm{x}$ given the noisy image $\bm{y}$ and noise variance $\sigma^2$.
It is noted that $\sigma$ is a scalar while $\bm{y}$ is an $H \times W \times C$ image ($C = 1$ for gray-scale image and $C = 3$ for color image).
To compensate the spatial dimension mismatch between $\sigma$ and $\bm{y}$, $\sigma$ is stretched into an $H \times W$ noise level map $\bm{M}$ with each element $\sigma$.
Consequently, the noise level map not only provides the conditional information for modulating convolutional activations of noisy image, but also offers a convenient means for handling spatially variant noise by simply setting $\bm{M}(i,j)$ as the local noise level at location $(i,j)$.

In general, a flexible non-blind image denoising model can be written as,
\begin{equation}\label{eqn_mf}
\bm{x} = \mathcal{F}(\bm{y},\bm{M}; \bm{\Theta}),
\end{equation}
where $\bm{\Theta}$ denotes the network parameters.
In \cite{zhang2018ffdnet,mildenhall2018burst,guo2019toward}, the network takes the concatenation of $\bm{y}$ and $\bm{M}$ as the input to deal with flexible denoising.
However, the above input concatenation disregards the heterogeneity of noise level map and noisy image.
As discussed in this section, the input concatenation can be treated as a channel-wise shifting modulation in the first convolution layer, thereby being limited in capturing the sophisticated interplay between $\bm{y}$ and $\bm{M}$.

In this section, we first analyze the limitation of input concatenation.
Then, we introduce the the residual shifting-based conditional feature modulation (RS-CFM) by considering convolution activations from $\bm{y}$ and $\bm{M}$ in a specific layer.
Finally, our CFMNet is given by incorporating multi-layer CFM with the U-Net backbone.

\begin{figure*}[!tbp]
\begin{center}
\begin{overpic}[width=1\textwidth]{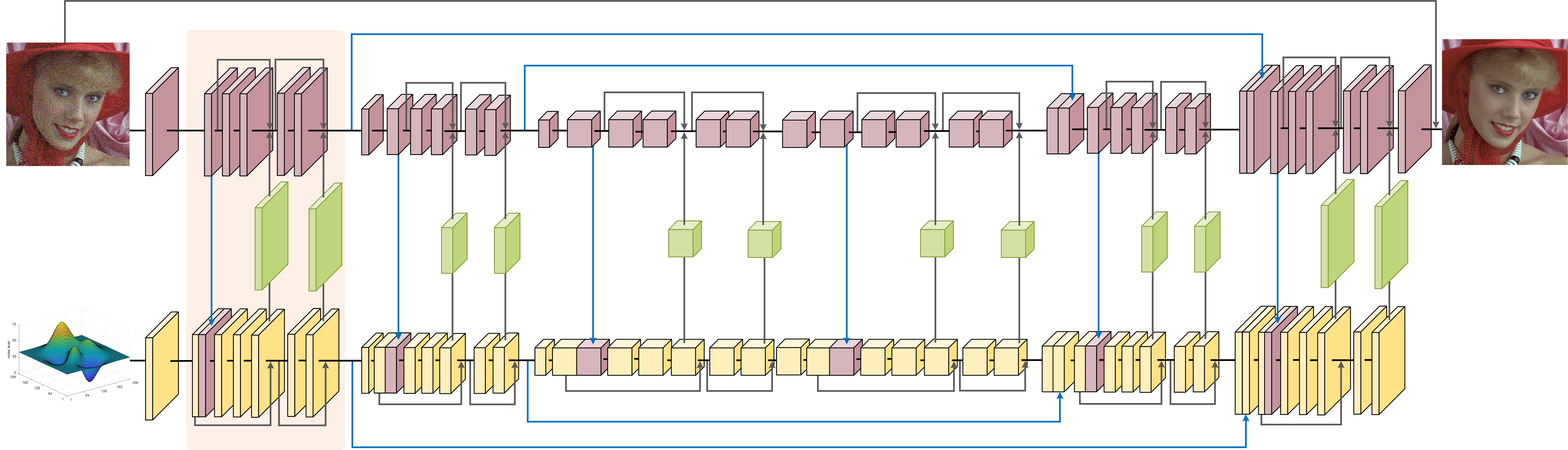} 
\put(1.6,17){\color{black}{\tiny Noisy Image}}
\put(1.2,2.0){\color{black}{\tiny Noise Level Map}}
\put(12.5,23.75){\color{black}{\tiny $\bm{f}^{l}$}}
\put(16,23.75){\color{black}{\tiny $\tilde{\bm{f}}^{l}$}}
\put(15.75,16){\color{black}{\tiny $\bm{s}^{l}$}}
\put(19.25,23.75){\color{black}{\tiny ${\hat{{\bm{f}}}}^{l}$}}
\put(18.75,16){\color{black}{\tiny ${\bm{s}'}^{l}$}}
\put(12.25,9.25){\color{black}{\tiny $\bm{g}^{l}$}}
\put(17.25,9.25){\color{black}{\tiny $\tilde{\bm{g}}^{l}$}}
\put(20.75,9.25){\color{black}{\tiny ${\hat{{\bm{g}}}}^{l}$}}
\put(17,25.5){\color{black}{\tiny 64$\times$64$\times$64}}
\put(14.5,25.15){\color{black}{\tiny Sum}}
\put(27,24){\color{black}{\tiny 128$\times$32$\times$32}}
\put(43,23.25){\color{black}{\tiny 256$\times$16$\times$16}}
\put(18,0.5){\color{black}{\tiny RS-CFM}}
\put(4.5,26.5){\color{black}{\tiny 3$\times$64$\times$64}}
\put(93.5,26.7){\color{black}{\tiny 3$\times$64$\times$64}}

\begin{turn}{270}
\put(-16.5,13.75){\color{black}{\tiny Concatenation}}
\end{turn}
\end{overpic}
\caption{Illustration of the network structure of our CFMNet. CFMNet consists of an image branch and a noise level map branch, each of which adopts an U-Net architecture.}\label{Framework}
\end{center}\vspace{-0.65cm}
\end{figure*}
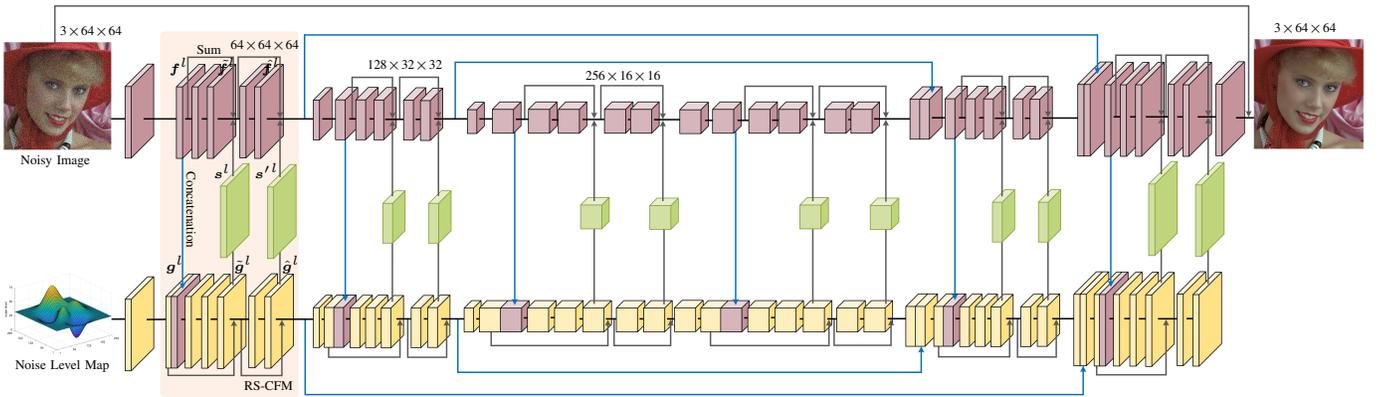
\label{sec:method}
\subsection{Revisiting FFDNet as Channel-wise Shifting Modulation}
In FFDNet, the network input is the concatenation $(\bm{y},\bm{M})$ with the dimension $H \times W \times (C+1)$.
The $k$-th channel of feature map in the first layer (i.e., $l=1$) can then be computed as,
\begin{equation}\label{eqn_FFD}
\bm{f}_k^{1} = \sum\nolimits_{c=1}^C \bm{w}_{c,k}^{1} \ast \bm{x}_c + \bm{w}_{C+1,k}^{1} \ast \bm{M} + {b}_k^{1},~~ k = 1, 2, \ldots , K_1,
\end{equation}
where $\ast$ denotes the convolution operation, and $K_1$ denotes the channel number of the first layer.
$\bm{w}_{c,k}^{1}$ is the convolution kernel for the $c$-th input map and the $k$-th output feature map, and ${b}_k^1$ denotes the $k$-th bias.

For spatially invariant AWGN, all the elements in $\bm{M}$ are of the same noise level $\sigma$.
Thus, all the elements in $\bm{w}_{C+1,k}^{1} \ast \bm{M}$ are also spatially invariant have the same value ${b}_k^{1,\bm{M}}$.
Then, Eqn.~(\ref{eqn_FFD}) can be equivalently reformulated as,
\begin{equation}\label{eqn_FFD_2}
\bm{f}_k^{1} = \sum\nolimits_{c=1}^C \bm{w}_{c,k}^{1} \ast \bm{x}_c + {{s}_k^{1}}(\sigma),~~ k = 1, 2, \ldots , K_1,
\end{equation}
where ${{s}_k^{1}}(\sigma) = {b}_k^{1,\bm{M}} + {b}_k^{1}$ denotes the channel-wise bias depending on the noise level $\sigma$.
From Eqn.~(\ref{eqn_FFD_2}), the input concatenation adopted in FFDNet can be treated as a normalization-free channel-wise shifting modulation conditioned on the noise level.
However, such modulation is only conducted in the first layer and the role of noise level is then diminished for the succeeding layers, making the input concatenation inadequate to capture the complex relation between $\bm{y}$ and $\bm{M}$.
%
Moreover, only the shifting modulation is considered in FFDNet. Thus, it is interesting to investigate the effect of feature modulation form and design a proper one.
Finally, as shown in Fig.~\ref{example}(b)(c), the effect of noise level may vary for different image contents (e.g., smooth region and detailed texture), which is also an interesting issue to be studied for balancing noise removal and detail preserving.

\subsection{Residual Shifting-based Conditional Feature Modulation}
%

%
In this subsection, we suggest a layer-wise residual shifting-based conditional feature modulation (RS-CFM) to overcome the limitations of input concatenation.
%
%
%
%
{Unlike affine transformation involving scaling and shifting, we empirically find that layer-wise residual shifting-based modulation works well and can be regarded as a multi-layer extension of input concatenation~\cite{zhang2018ffdnet}.
Moreover, in input concatenation the shifting bias depends only on the noise level, while in RS-CFM the shifting map is determined by both noisy image and noise level map and thus can be spatially variant even for spatially invariant AWGN.
Without loss of generality, we use the convolutional activations in the $l$-th layer to explain our RS-CFM.
Denote by $\bm{f}^{l}$ and $\bm{g}^{l}$ the $l$-th layer of convolutional activations from the noisy image and noise level map, respectively.
As illustrated in Fig.~\ref{Framework}, $\bm{f}^{l}$ and $\bm{g}^{l}$ are concatenated as the input of {three} convolutional layers to generate $\tilde{\bm{g}}^{l}$.  
Then, a convolution layer is exploited to generate conditional shifting map $\bm{s}^{l}$ which takes $\tilde{\bm{g}}^{l}$ as the input.
Besides, we further use {two} convolution layers on $\bm{f}^{l}$ to generate $\tilde{\bm{f}}^{l}$.
With $\bm{f}^{l}$, $\tilde{\bm{f}}^{l}$ and $\bm{s}^{l}$, we introduce the first residual shifting operation as,
\begin{equation}\label{eqn:RS-1}
{\bm{f}^{\prime}}^{l} = \bm{f}^{l} + (\tilde{\bm{f}}^{l}+ \bm{s}^{l}),
\end{equation}
where ${\bm{f}^{\prime}}^{l}$ is the output of the residual shifting operation.
Analogously, the conditional feature ${\bm{g}^{\prime}}^{l}$ is updated in a residual learning manner,
\begin{equation}\label{eqn:RS-g}
{\bm{g}^{\prime}}^{l} = \bm{g}^{l} + \tilde{\bm{g}}^{l}.
\end{equation}
One can easily see that the above equation is a generalization of the input concatenation in Eqn.~(\ref{eqn_FFD_2}) to intermediate layers.
We further introduce another residual shifting block for modulating ${\bm{f}^{\prime}}^{l}$ conditioned on ${\bm{g}^{\prime}}^{l}$.
Analogous to the first residual shifting block, we use two convolution layers on ${\bm{g}^{\prime}}^{l}$ to generate ${\hat{{\bm{g}}}}^{l}$.
And a convolution layer is also deployed on ${\hat{{\bm{g}}}}^{l}$ to generate the shifting map ${\bm{s}'}^{l}$.
After updating the convolutional feature ${\hat{{\bm{f}}}}^{l}$ from ${\bm{f}^{\prime}}^{l}$ with {two} convolution layers, the convolutional activations for the succeeding layer can be obtained by,
\begin{equation}\label{eqn:RS-2}
{\bm{f}''}^{l} = {\bm{f}^{\prime}}^{l} + ({\hat{{\bm{f}}}}^{l}+ {\bm{s}'}^{l}),
\end{equation}
\begin{equation}\label{eqn:RS-g2}
{\bm{g}''}^{l} = {\bm{g}^{\prime}}^{l} + {\hat{{\bm{g}}}}^{l}.
\end{equation}

It is noteworthy that each RS-CFM module involves two residual shifting operations.
Obviously, it is feasible to utilize more residual shifting operations, but we empirically find that two residual shifting operations generally work well in balancing performance and efficiency.
%
%
Moreover, affine transformation including scaling and shifting has been widely adopted in most existing conditional modulation methods~\cite{Perez2017FiLM, wang2018recovering}.
When ${ {\tilde{\bm{f}}^{l}} \approx 0}$, residual shifting can also be treated as a special case of affine transformation and an extension of channel-wise shifting modulation adopted in the first layer of FFDNet.
Instead, it is also feasible to directly learn $({\tilde{\bm{f}}^{l}} + {\bm{s}}^{l})$.
But we empirically find that RS-CFM is beneficial to the training stability and denoising performance in comparison to conventional shifting modulation.
As for existing flexible denoising networks such as FFDNet~\cite{zhang2018ffdnet}, the noise level map is only employed to perform feature modulation in the first layer, and its role is then diminished in the succeeding layers.
In comparison, our RS-CFM can be incorporated with existing denoising networks and be deployed in multiple layers to better leverage noise level map.

Besides, the effect of noise level also depends on the image content.
For example, the denoising result is robust for smooth region when the input noise level is larger than the ground-truth one.
So it is possible to utilize larger input noise level to suppress the artifact of the denoising result in smooth regions.
In contrast, the denoising result may be over-smoothing for small-scale texture region when the input noise level is matched with or larger than the ground-truth one.
Nonetheless, more fine details can be retained by adopting slightly smaller input noise level.
To sum up, it is suggested to adapt the input noise level to image content, which explain that our RS-CFM takes the convolutional activations from both noisy image and noise level map to generate the shifting map.
Consequently, the conditional shifting map by our RS-CFM is spatially variant.
Also, instead of handcrafted tuning on noise level map according to local image content, our residual shifting map is end-to-end learnable from the training data and model objective.

It is noted that spatially variant conditional modulation has also been adopted in SPADE~\cite{park2019semantic} and SFT~\cite{wang2018recovering}.
While SPADE~\cite{park2019semantic} is a conditional normalization method, SFT~\cite{wang2018recovering} and our RS-CFM are normalization-free conditional modulation.
SPADE~\cite{park2019semantic} and SFT~\cite{wang2018recovering} exploit semantic labeling as the conditional information for image synthesis and SISR, and adopt affine transformation for feature modulation.
In contrast, we utilize noise level map as the conditional information and and suggest the residual shifting-based modulation for flexible non-blind image denoising.
Finally, the spatial variation of conditional modulation in~\cite{wang2018recovering,park2019semantic} comes from the spatial layout of semantic labeling and is independent with the feature to be modulated.
Instead, our RS-CFM considers the convolutional activations from both noisy image and noise level map to generate the conditional shifting map.
That is, even for spatially invariant noise level map, our RS-CFM can produce spatially variant shifting map for better tradeoff between noise removal and detail preserving.

\begin{table*}[htbp]\large
\centering
\caption{PSNR/SSIM results of different  methods for gray-scale image denoising on Set12 dataset with noise standard deviation 15, 25, 50 and 75.}
\label{gray_psnr}
\resizebox{\textwidth}{!}{
\begin{tabular}{|c|c|c|c|c|c|c|c|c|c|c|c|}
\hline

Datasets                   & $\sigma$ & BM3D~\cite{dabov2007image}& DnCNN~\cite{zhang2017beyond}& FFDNet~\cite{zhang2018ffdnet}& MemNet~\cite{tai2017memnet}&N3Net~\cite{Pl2018Neural}& NLRN~\cite{liu2018non}& FOCNet~\cite{jia2019focnet}& DHDN~\cite{park2019densely}& MWCNN~\cite{liu2018multi}& CFMNet   \\ \hline
                           & 15    & 32.37/0.8952 & 32.86/0.9027 & 32.79/0.9062 &-             & -            & 33.16/0.9099& 33.07/- & -            &33.15/0.9088  & \textbf{33.16/0.9113} \\
                           & 25    & 29.97/0.8505 & 30.44/0.8601 & 30.45/0.8662 &-             & 30.53/0.8668 & 30.79/0.8704& 30.73/- &-             & 30.79/0.8711 & \textbf{30.86/0.8741} \\
                           & 50    & 26.72/0.7676 & 27.18/0.7827 & 27.34/0.7890 & 27.38/0.7931 & 27.44/0.7931 & 27.64/0.7959& 27.68/- & 27.66/0.8030 & 27.74/0.8048 & \textbf{27.80/0.8053} \\
\multirow{-4}{*}{Set12}    & 75    & 24.87/0.7044 & 25.20/0.7184 &25.52/0.7298  & -            & -            & -           &       - & -            & 26.01/\textbf{0.7556} & \textbf{26.02}/0.7527\\ \hline
                           & 15    & 31.08/0.8722 & 31.73/0.8906 & 31.63/0.8957 & -            & -            & 31.88/0.8983& 31.83/- &-             & 31.86/0.8947 & \textbf{31.88/0.9003} \\
                           & 25    & 28.57/0.8017 & 29.23/0.8148 & 29.21/0.8352 & -            & 29.30/0.8377 & 29.40/0.8383& 29.38/- & -            & 29.41/0.8361 & \textbf{29.45/0.8431} \\
                           & 50    & 25.62/0.6869 & 26.23/0.7189 & 26.32/0.7283 & 26.35/0.7294 & 26.39/0.7321 & 26.45/0.7313& 26.50/- & 26.30/0.7263 & 26.54/0.7364 & \textbf{26.56/0.7412} \\
\multirow{-4}{*}{BSD68}    & 75    & 24.20/0.6216 & 24.64/0.6465 & 24.81/0.6564 & -            & -            & -           &       - & -            & 25.07/0.6744 & \textbf{25.07/0.6745}\\ \hline
                           & 15    & 32.37/0.8952 & 32.67/0.9250 &32.49/0.9303  & -            & -            & 33.45/0.9354& 33.15/- & -            & 33.17/0.9357 & \textbf{33.18/0.9378} \\
                           & 25    & 29.97/0.8505 & 29.97/0.8792 & 30.03/0.8928 & -            & 30.19/0.8926 & 30.94/0.9018& 30.64/- & -            & 30.66/0.9026 & \textbf{30.79/0.9063} \\
                           & 50    & 26.72/0.7676 & 26.28/0.7869 & 26.65/0.8102 &26.64/0.8024  & 26.82/0.8148 & 27.49/0.8279& 27.40/- & 27.66/0.8030 &27.42/0.8371  &\textbf{27.56/0.8396} \\
\multirow{-4}{*}{Urban100} & 75    & 23.93/0.7022 & 23.99/0.7055 & 24.62/0.7380 & -            & -            & -           &       - & -            & 25.68/0.7858 &\textbf{25.98/0.7933}\\ \hline
\end{tabular}%
}
\end{table*}
\subsection{Architecture of CFMNet}
In this subsection, we introduce the network structure of our CFMNet by incorporating a backbone U-Net~\cite{ronneberger2015u} with multi-layer RS-CFM.
The reason to choose the U-Net architecture is that it is promising in denoising performance and has been adopted in several state-of-the-art deep denoising models~\cite{Mao2016Image,ronneberger2015u,liu2018multi,guo2019toward}.
A typical U-Net generally involves an encoder and a decoder subnet.
In particular, pooling and up-convolution are respectively introduced in the encoder and decoder, which can enlarge the receptive field and is expected to benefit denoising performance~\cite{ronneberger2015u}.
Moreover, the features for an encoder layer is concatenated with the decoder feature from the corresponding layer in a symmetric manner, which is also helpful in easing the network training~\cite{ronneberger2015u}.

Fig.~\ref{Framework} illustrates the network structure of our CFMNet, which consists of an image branch and a noise level map branch.
Each branch adopts an U-Net architecture with two $2 \times 2$ max-pooling layers for downsampling and two $2 \times 2$ transposed convolution layers for upsampling. The number of feature maps in different scales is set to 64, 128 and 256, respectively.
Two RS-CFM modules are added before each downsampling or after each upsampling operator.
The size of all convolution kernels in RS-CFM is $3 \times 3$.
ReLU nonlinearity is adopted for all convolution (Conv) layers except for the last ones and the Conv layers deployed to generate shifting maps.
{For faster convergence, batch normalization (BN) is employed but is then merged into the previous Conv layer at the end of network training.}
%
The encoder features from each branch are concatenated with the corresponding decoder features.
Instead of directly learning the denoising result $\hat{\bm{x}} = \mathcal{F}(\bm{y}, \bm{M}; \bm{\Theta})$, we leverage the residual learning formulation~\cite{zhang2017beyond} to predict the denoising result by $\hat{\bm{x}} = \bm{y} + \mathcal{R}(\bm{y},\bm{M}; \bm{\Theta})$.

\subsection{Model Objective and Learning}
Our CFMNet takes both the noise level map $\bm{M}$ and noisy image $\bm{y}$ as the input, and adopts residual learning to predict the residual between the ground-truth clean image and noisy image.
Denote by $\{(\bm{y}_i, \bm{M}_i, \bm{x}_i)\}_{i=1}^{N}$ a training set, where $\bm{y}_i$, $\bm{M}_i$ and $\bm{x}_i$ denote the $i$-th noisy image, noise level map, and ground-truth clean image, respectively.
Our CFMNet can then be learned by minimizing the mean squared error (MSE) loss,
\begin{equation}\label{eq:loss}
  \mathcal{L}(\bm{\Theta}) = \frac{1}{2N}\sum\nolimits_{i=1}^N\| {\bm{x}}_i - ({\bm{y}}_i - \mathcal{R}(\bm{y},\bm{M}; \bm{\Theta}))  \|^2,
\end{equation}
where $\bm{\Theta}$ denotes the network parameters to be learned.

The ADAM optimizer~\cite{kingma2014adam} with the default setting is utilized to learn our CFMNet.
We adopt the initialization method~\cite{he2015delving}.
{The learning rate begins with $1e{-4}$ and then exponentially decays to $1e{-6}$ in 75 epochs. Then, we merge the parameters of batch normalization into the parameters in the previous Conv layer. Finally, the learning rate decays to from $1e{-6}$ to $1e{-7}$ for 10 epochs to fine-tune our CFMNet.}

\section{Experiments}
Extensive experiments are conducted to evaluate our CFMNet for gray-scale and color image denoising.
In particular, we test the effect of residual shifting, the role of image feature in modulation, the numbers of residual shifting operations in RS-CFM and compare our CFMNet with the state-of-the-art gray-scale and color image denoising methods.
All the source code, pre-trained models will be publicly available at
{\url{https://github.com/dujiazhi/CFMNet}}.

\label{sec:experiments}
\subsection{Experimental Setting}
In this subsection, we introduce the training and testing sets used in our experiments, and describe the implementation details for training our CFMNet models.
Following~\cite{zhang2018ffdnet}, our training set is constituted by 400 images from the Berkeley Segmentation Dataset (BSD) dataset~\cite{MartinFTM01}, 400 images from the ImageNet validation set~\cite{deng2009imagenet}, and 4,744 images from the Waterloo Exploration Database~\cite{ma2016gmad}.
The gray-scale and color images in the training set are respectively used to train the gray-scale and color image denoising models.
For evaluating gray-scale image denoising models, we adopt three testing datasets including (i) the Set12 dataset consisting of 12 widely-used testing images~\cite{zhang2017beyond}, (ii) the BSD68 dataset containing 68 images from the BSD test set~\cite{roth2009fields}, and (iii) the Urban100 dataset involving 100 high-quality images with rich real-world structures~\cite{huang2015single}.
Moreover, three color testing datasets are used to evaluate the color image denoising models, i.e., the CBSD68 dataset~\cite{roth2009fields} as the color version of BSD68, the Kodak24 dataset of 24 centrally cropped $500 \times 500$ images from Kodak PhotoCD~\cite{franzen1999kodak}, and the McMaster dataset of 18 cropped $500 \times 500$ images~\cite{zhang2011color}.

The training of CFMNet requires a set of triplets $\{(\bm{y}_i, \bm{M}_i, \bm{x}_i)\}_{i=1}^{N}$.
To this end, we randomly crop $N = 64\times 4,038$ patches of size $64 \times 64$ from the images in the training set in each epoch.
For each patch, we randomly sample a noise standard deviation $\sigma_i \in [0, 75]$ to synthesize the noisy patch.
Then, the noise standard deviation $\sigma_i$ is stretched into an $64 \times 64$ noise level map $\bm{M}_i$.
We note that our CFMNet is suggested to handle AWGN with various noise levels or even spatially variant Gaussian noise, and thus it is expected to employ spatially variant $\bm{M}_i$ during training.
Fortunately, our CFMNet is a fully convolutional network (FCN) trained on image patches.
That is, the denoised result at a spatial position is determined only by its local noisy input and local noise level.
Consequently, albeit we train our CFMNet using spatially invariant $\bm{M}_i$ with various noise levels, the learned model can be directly applied to handle spatially variant Gaussian noise.
Thus, only spatially invariant $\bm{M}_i$ is considered when training our CFMNet.
Standard data augmentation methods, e.g., random flipping, rotation and scaling, are used in the training procedure.
All the experiments are conducted in the PyTorch environment on a PC equipped with Intel(R) Xeon(R) E3-1231 v3 CPU 3.40GHz and a GeForce GTX 2080Ti GPU.
With the above experimental setting, we require about two days to train a CFMNet model.
\begin{figure*}[!htbp]
\tiny
\small
	\newlength\fsfourteen
	\setlength{\fsfourteen}{-4.1mm}
	\scriptsize
	\centering
	\begin{tabular}{cc}
	\tiny
	\scriptsize
	\footnotesize
	\small
		\hspace{-0.3cm}
		\begin{adjustbox}{valign=t}
			\begin{tabular}{ccccc}
				\includegraphics[width=0.19\textwidth]{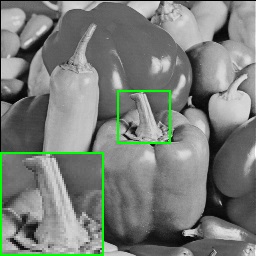} \hspace{\fsfourteen} &
				\includegraphics[width=0.19\textwidth]{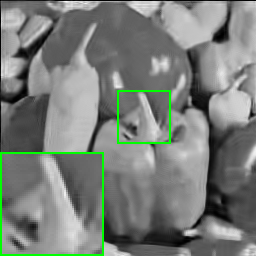}  \hspace{\fsfourteen} &
				\includegraphics[width=0.19\textwidth]{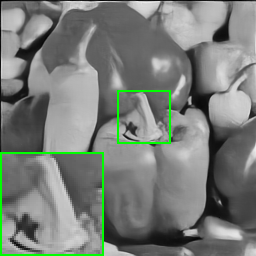} \hspace{\fsfourteen} &
				\includegraphics[width=0.19\textwidth]{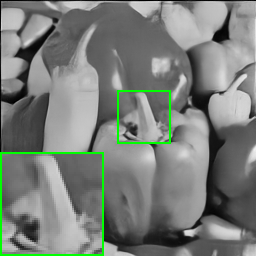} \hspace{\fsfourteen} &
				\includegraphics[width=0.19\textwidth]{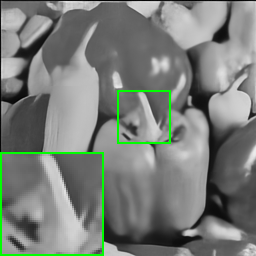} 
				\\
				(a)\hspace{\fsfourteen} &
				(b)\hspace{\fsfourteen} &
				(c)\hspace{\fsfourteen} &
				(d)&
				(e)\hspace{\fsfourteen}
				\\
				\includegraphics[width=0.19\textwidth]{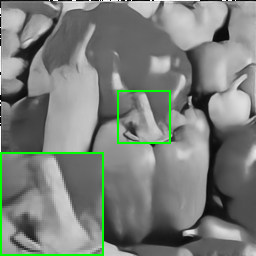} \hspace{\fsfourteen} &
				\includegraphics[width=0.19\textwidth]{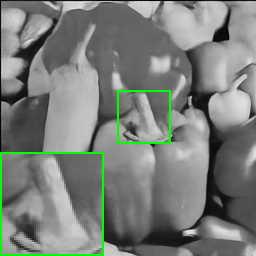} \hspace{\fsfourteen} &
				\includegraphics[width=0.19\textwidth]{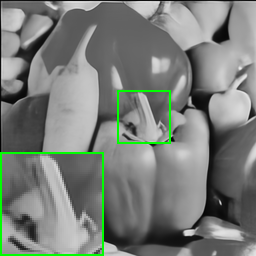} \hspace{\fsfourteen} &
				\includegraphics[width=0.19\textwidth]{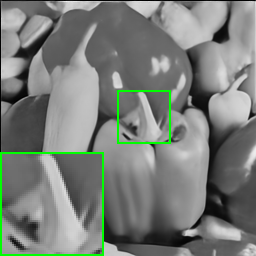} \hspace{\fsfourteen} &
				\includegraphics[width=0.19\textwidth]{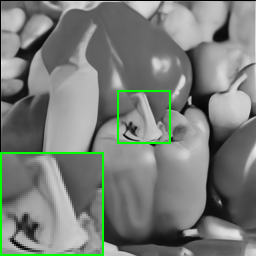}
				\\
				(f)\hspace{\fsfourteen} &
				(g)\hspace{\fsfourteen} &
				(h)\hspace{\fsfourteen} &
				(i)\hspace{\fsfourteen} &
				(j)
				\\
			\end{tabular}
		\end{adjustbox}
	\end{tabular}
\caption{Denoising results by different methods on image \emph{pepper} from Set12 with $\sigma = 50$. (a) Ground-truth; (b) BM3D; (c) DnCNN; (d) FFDNet; (e) MemNet; (f) N3Net; (g) NLRN; (h) DHDN; (i) MWCNN; (j) CFMNet (Ours).}\label{figset12_1}
\end{figure*}

\begin{figure*}[!htbp]
\tiny
\small
	\setlength{\fsfourteen}{-4.1mm}
	\scriptsize
	\centering
	\begin{tabular}{cc}
	\tiny
	\scriptsize
	\footnotesize
	\small
		\hspace{-0.3cm}
		\begin{adjustbox}{valign=t}
			\begin{tabular}{ccccc}
				\includegraphics[width=0.19\textwidth]{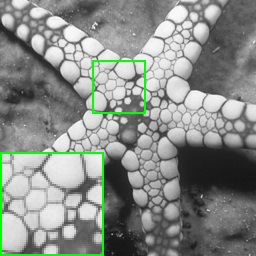} \hspace{\fsfourteen} &
				\includegraphics[width=0.19\textwidth]{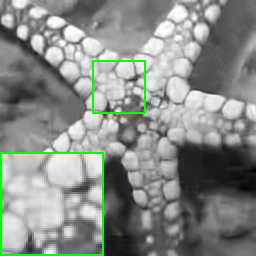}  \hspace{\fsfourteen} &
				\includegraphics[width=0.19\textwidth]{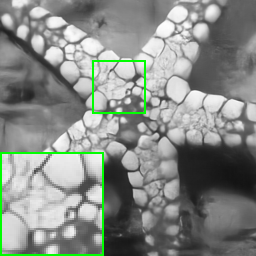} \hspace{\fsfourteen} &
				\includegraphics[width=0.19\textwidth]{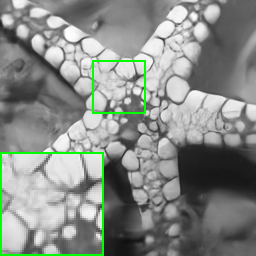} \hspace{\fsfourteen} &
				\includegraphics[width=0.19\textwidth]{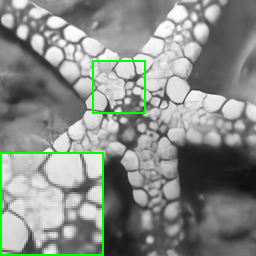} 
				\\
				(a)\hspace{\fsfourteen} &
				(b)\hspace{\fsfourteen} &
				(c)\hspace{\fsfourteen} &
				(d)&
				(e)\hspace{\fsfourteen}
				\\
				\includegraphics[width=0.19\textwidth]{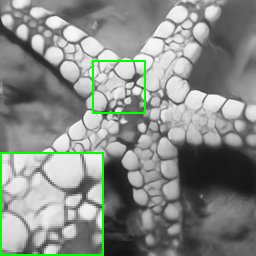} \hspace{\fsfourteen} &
				\includegraphics[width=0.19\textwidth]{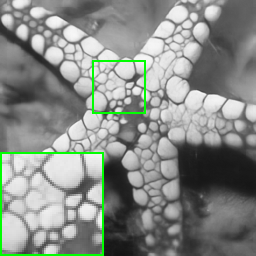} \hspace{\fsfourteen} &
				\includegraphics[width=0.19\textwidth]{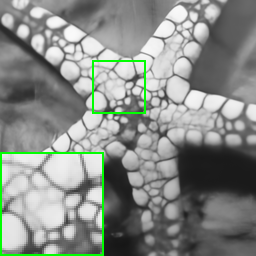} \hspace{\fsfourteen} &
				\includegraphics[width=0.19\textwidth]{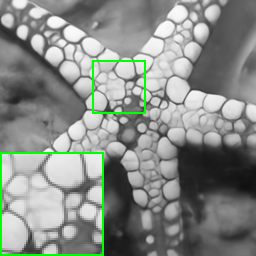} \hspace{\fsfourteen} &
				\includegraphics[width=0.19\textwidth]{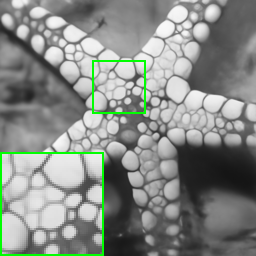}
				\\
				(f)\hspace{\fsfourteen} &
				(g)\hspace{\fsfourteen} &
				(h)\hspace{\fsfourteen} &
				(i)\hspace{\fsfourteen} &
				(j)
				\\
			\end{tabular}
		\end{adjustbox}
	\end{tabular}
\caption{Denoising results by different methods on image \emph{starfish} from Set12 with $\sigma = 50$. (a) Ground-truth; (b) BM3D; (c) DnCNN; (d) FFDNet; (e) MemNet; (f) N3Net; (g) NLRN; (h) DHDN; (i) MWCNN; (j) CFMNet (Ours).}\label{figset12_2}
\end{figure*}

\begin{figure*}[!htbp]
\tiny
\small
	\setlength{\fsfourteen}{-4.1mm}
	\scriptsize
	\centering
	\begin{tabular}{cc}
	\tiny
	\scriptsize
	\footnotesize
	\small
		\hspace{-0.3cm}
		\begin{adjustbox}{valign=t}
			\begin{tabular}{ccccc}
				\includegraphics[width=0.19\textwidth]{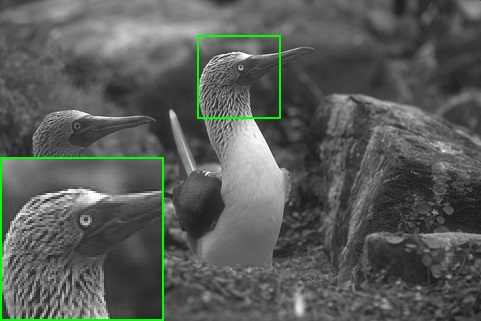} \hspace{\fsfourteen} &
				\includegraphics[width=0.19\textwidth]{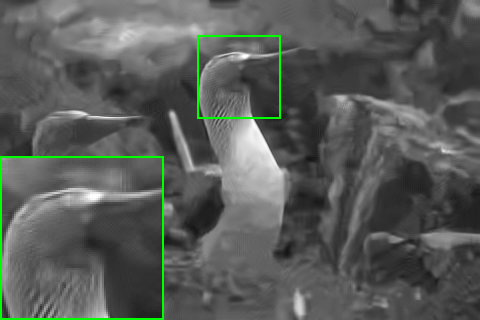}  \hspace{\fsfourteen} &
				\includegraphics[width=0.19\textwidth]{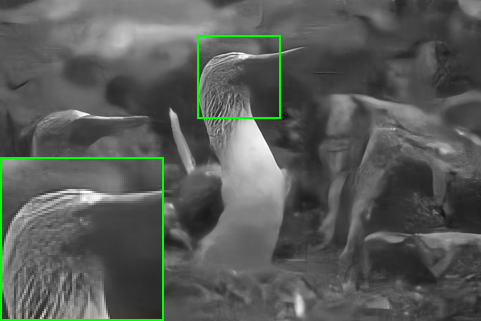} \hspace{\fsfourteen} &
				\includegraphics[width=0.19\textwidth]{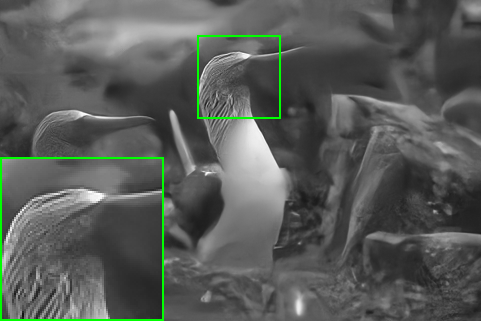} \hspace{\fsfourteen} &
				\includegraphics[width=0.19\textwidth]{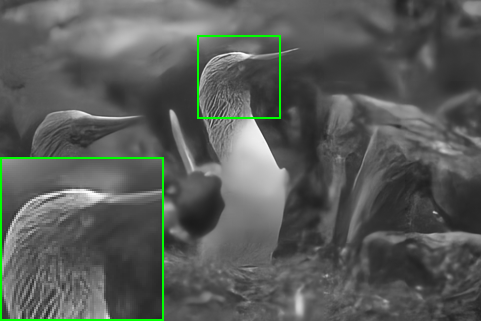}
				\\
				(a)\hspace{\fsfourteen} &
				(b)\hspace{\fsfourteen} &
				(c)\hspace{\fsfourteen} &
				(d)&
				(e)\hspace{\fsfourteen}
				\\
				\includegraphics[width=0.19\textwidth]{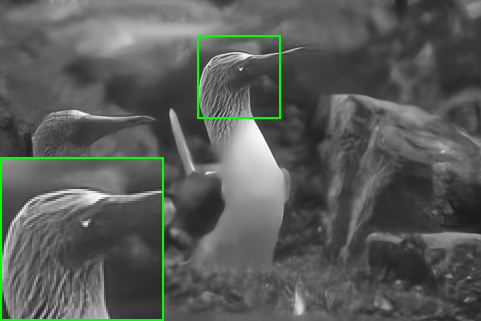} \hspace{\fsfourteen} &
				\includegraphics[width=0.19\textwidth]{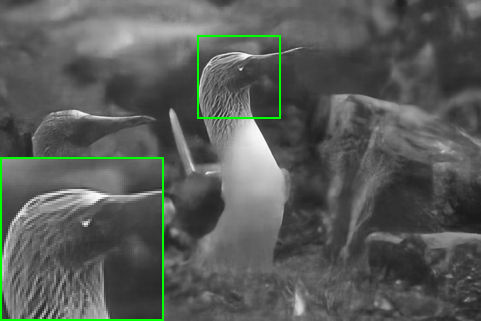} \hspace{\fsfourteen} &
				\includegraphics[width=0.19\textwidth]{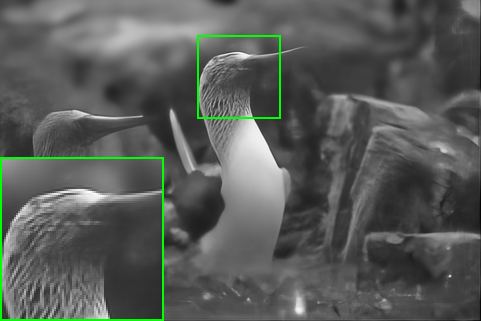} \hspace{\fsfourteen} &
				\includegraphics[width=0.19\textwidth]{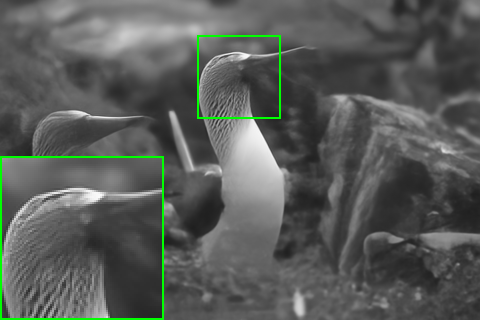} \hspace{\fsfourteen} &
				\includegraphics[width=0.19\textwidth]{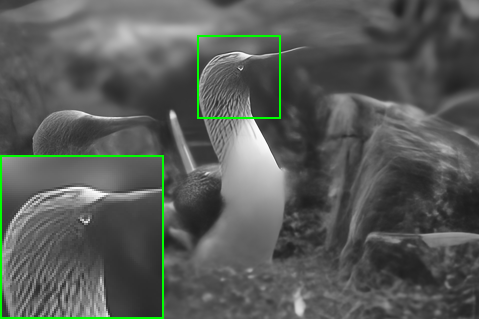}
				\\
				(f)\hspace{\fsfourteen} &
				(g)\hspace{\fsfourteen} &
				(h)\hspace{\fsfourteen} &
				(i)\hspace{\fsfourteen} &
				(j)
				\\
			\end{tabular}
		\end{adjustbox}

	\end{tabular}
\caption{Denoising results by different methods on image \emph{test004} from the BSD68 dataset with $\sigma=50$. (a) Ground-truth; (b) BM3D; (c) DnCNN; (d) FFDNet; (e) MemNet; (f) N3Net; (g) NLRN; (h) DHDN; (i) MWCNN; (j) CFMNet (Ours). }\label{figset68_1}
\end{figure*}

\begin{table*}[htbp]\footnotesize
\centering
\caption{PSNR/SSIM results by different methods for color image denoising on four testing datasets with noise standard deviation 15, 25, 50, and 75.}
\label{color_psnr}
\begin{tabular}{|c|c|c|c|c|c|c|c|c|}
\hline

Datasets                   & $\sigma$ & CBM3D~\cite{dabov2007image}& CDnCNN~\cite{zhang2017beyond} & FFDNet~\cite{zhang2018ffdnet}&  DHDN~\cite{park2019densely}& CMWCNN~\cite{liu2018multi}& CFMNet        \\ \hline
                           & 15    & 33.50/0.9215 & 33.98/0.9303 & 33.81/0.9318 & -/-          & 34.08/0.9316 & \textbf{34.26/0.9376}          \\
                           & 25    & 30.69/0.8672 & 31.31/0.8848 & 31.20/0.8874 & -/-          & 31.40/0.8873 & \textbf{31.64/0.8969}         \\
                           & 50    & 27.36/0.7626 & 28.01/0.7925 & 28.00/0.7977 & 28.05/0.7951 & 28.26/0.8031 & \textbf{28.46/0.8149}          \\
\multirow{-4}{*}{CBSD68}   & 75    & 25.73/0.6957 & 26.31/0.7279 & 26.30/0.7293 & -/-          & 26.60/0.7403 & \textbf{26.79/0.7532}          \\ \hline
                           & 15    & 34.26/0.9147 & 34.73/0.9224 & 34.58/0.9230 & -/-          & 34.88/0.9250 & \textbf{35.09/0.9301}          \\
                           & 25    & 31.67/0.8670 & 32.24/0.8800 & 32.15/0.8812 & -/-          & 32.38/0.8845 & \textbf{32.69/0.8927}          \\
                           & 50    & 28.44/0.7760 & 29.03/0.7976 & 29.03/0.7987 & 29.40/0.8123 & 29.37/0.8115 & \textbf{29.64/0.8195}         \\
\multirow{-4}{*}{Kodak24}  & 75    & 26.82/0.7184 & 27.28/0.7395 & 27.33/0.7380 & -/-          & 27.72/0.7573 & \textbf{27.98/0.7656}          \\ \hline
                           & 15    & 34.03/0.9114 & 34.81/0.9228 & 34.56/0.9239 & -/-          & 34.92/0.9261 & \textbf{35.23/0.9336}          \\
                           & 25    & 31.63/0.8699 & 32.48/0.8889 & 32.35/0.8907 & -/-          & 32.55/0.8924 & \textbf{32.98/0.9043}          \\
                           & 50    & 28.48/0.7911 & 29.22/0.8175 & 29.27/0.8217 & 29.59/0.8248 & 29.61/0.8323 & \textbf{29.99/0.8453}          \\
\multirow{-4}{*}{McMaster} & 75    & 26.76/0.7351 & 27.12/0.7553 & 27.43/0.7652 & -/-          & 27.90/0.7854 & \textbf{28.23/0.7987}          \\ \hline
                           & 15    & 33.93/0.9408 & 34.11/0.9436 & 33.80/0.9443 & -/-          & 34.20/0.9456 & \textbf{34.57/0.9512}          \\
                           & 25    & 31.36/0.9092 & 31.66/0.9145 & 31.45/0.9163 & -/-          & 31.68/0.9173 & \textbf{32.33/0.9282}          \\
                           & 50    & 27.93/0.8404 & 28.16/0.8490 & 28.17/0.8538 & 29.26/0.8759 & 28.59/0.8636 & \textbf{29.30/0.8804}          \\
\multirow{-4}{*}{Urban100} & 75    & 25.95/0.7824 & 25.96/0.7844 & 26.18/0.7965 & -/-          & 26.80/0.8177 & \textbf{27.47/0.8384}          \\ \hline
\end{tabular}%
\end{table*}

\subsection{Experiments on Gray-scale Image Denoising}
We compare our CFMNet with one traditional denoising method, i.e., BM3D~\cite{dabov2007image}, six CNN-based methods, i.e., DnCNN~\cite{zhang2017beyond}, FFDNet~\cite{zhang2018ffdnet}, MemNet~\cite{tai2017memnet}, FOCNet~\cite{jia2019focnet}, DHDN~\cite{park2019densely} and MWCNN~\cite{liu2018multi}, and two non-local networks, i.e., N3Net~\cite{Pl2018Neural} and
NLRN~\cite{liu2018non}.
On gray-scale image denoising, we consider spatially invariant AWGN with noise standard deviation 15, 25, 50 and 75.
Due to that the source code of FOCNet is not available, we simply adopt the results from the original paper~\cite{jia2019focnet}.

Table~\ref{gray_psnr} lists the average PSNR and SSIM results for gray-scale image denoising of different methods on three datasets, i.e., Set12, BSD68 and Urban100.
From Table~\ref{gray_psnr}, it can be seen that our CFMNet achieves the best PSNR results on all noise variance and all testing datasets.
%
%
Taking $\sigma=50$ and Set12 as an example, CFMNet outperforms the traditional model-based method BM3D~\cite{dabov2007image} by a large margin, i.e., $\sim$1.1dB.
In comparison with DnCNN~\cite{zhang2017beyond}, FFDNet~\cite{zhang2018ffdnet} and MemNet~\cite{tai2017memnet}, the improvements by our CFMNet can be about 0.5dB.
CFMNet does not require the time consuming non-local operations, and outperforms the non-local-based denoising networks N3Net~\cite{Pl2018Neural} and
NLRN~\cite{liu2018non} by more than 0.2dB.
Compared with the latest CNN denoisers such as MWCNN~\cite{liu2018multi}, FOCNet~\cite{jia2019focnet} and DHDN~\cite{park2019densely}, CFMNet can still achieve a PSNR gain of 0.1dB.
In terms of PSNR and SSIM, CFMNet outperforms the competing methods on all the test sets and noise variance except the SSIM result of MWCNN~\cite{liu2018multi} (0.7556) is slightly higher than that of CFMNet (0.7527) for $\sigma=75$ on Set12.
The result show that CFMNet is effective in flexible non-blind image denoising and performs
favorably against the state-of-the-art methods.
%

Visual comparison of denoising results is further provided to evaluate the denoising methods.
Using AWGN with $\sigma =50$ as an example, Figs.~\ref{figset12_1},~\ref{figset12_2} and~\ref{figset68_1} show the denoising results by different methods on three images from Set12 and BSD68.
All the competing methods except FOCNet~\cite{jia2019focnet} are considered in visual comparison because the source code and pre-trained models of FOCNet are publicly unavailable.
%
%
%
One can see that our CFMNet is more effective in restoring fine-scale textures and details, e.g., the shape of the stalk of image \emph{pepper} in Fig.~\ref{figset12_1}, the texture region of image \emph{starfish} in Fig.~\ref{figset12_2}, the eyeball of image \emph{test004} in Fig.~\ref{figset68_1}.
In comparison to the competing methods, our CFMNet is promising in balancing noise removal and detail preserving, which can be ascribed to (i) considering both noisy image and noise level map in feature modulation and (ii) deploying CFM in multiple layers.



\begin{figure*}[!htbp]
\tiny
\small
	\scriptsize
	\centering
	\begin{tabular}{cc}
	\tiny
	\scriptsize
	\footnotesize
	\small
		\hspace{-0.4cm}
		\begin{adjustbox}{valign=t}
			\begin{tabular}{cccc}
                \includegraphics[width=0.24\textwidth]{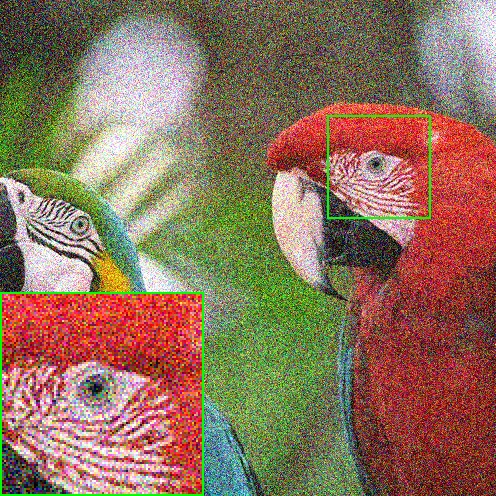}&\hspace{-4.2mm}
                \includegraphics[width=0.24\textwidth]{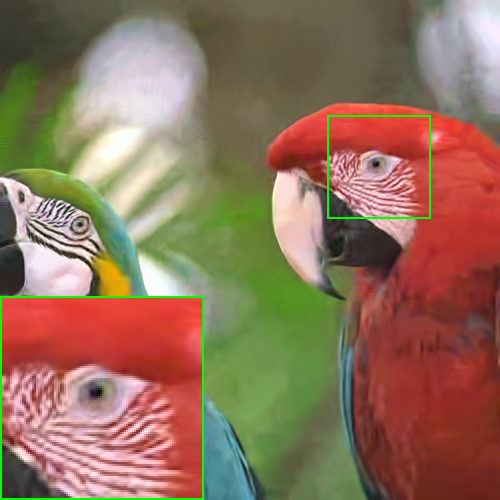}&\hspace{-4.2mm}
                \includegraphics[width=0.24\textwidth]{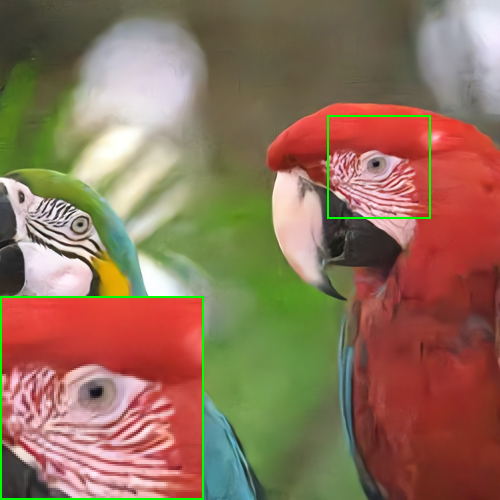}&\hspace{-4.2mm}
                \includegraphics[width=0.24\textwidth]{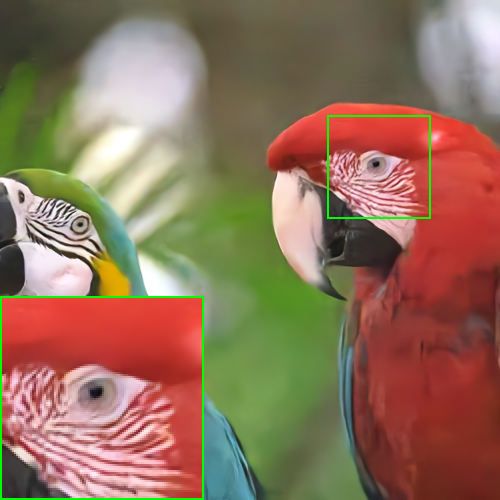}
				\\
				(a)&\hspace{-4.2mm} 			
				(b)& \hspace{-4.2mm}
				(c)& \hspace{-4.2mm}
				(d)
				\\
                \includegraphics[width=0.24\textwidth]{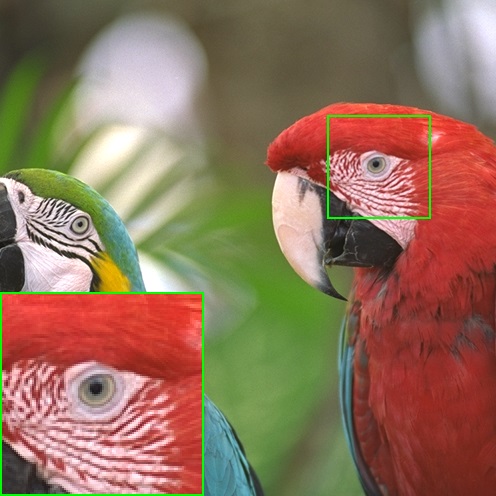}&\hspace{-4.2mm}
                \includegraphics[width=0.24\textwidth]{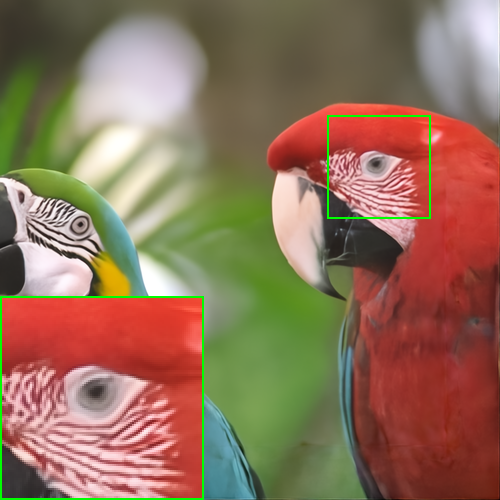}&\hspace{-4.2mm}
                \includegraphics[width=0.24\textwidth]{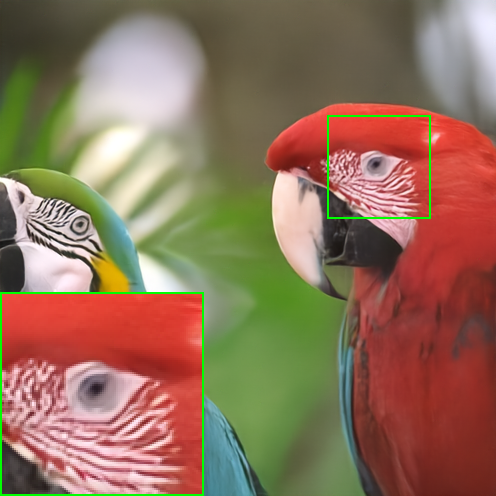}&\hspace{-4.2mm}
                \includegraphics[width=0.24\textwidth]{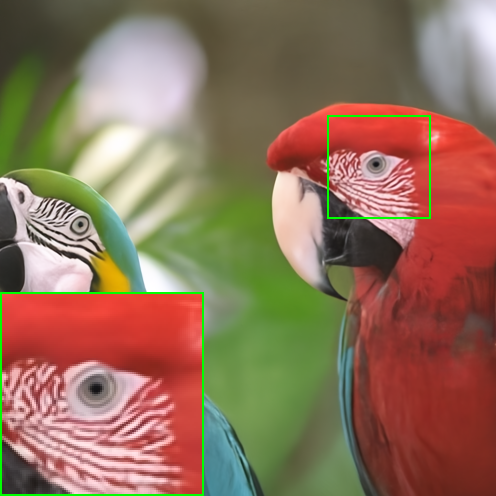}
				\\
                (e)&\hspace{-4.2mm}
				(f)&\hspace{-4.2mm}
				(g)& \hspace{-4.2mm}
				(h)
				\\
			\end{tabular}
		\end{adjustbox}

	\end{tabular}
\caption{Denoising results by different methods on image \emph{023} from the Kodak24 dataset with $\sigma = 50$. (a) Noisy image; (b) BM3D; (c) DnCNN; (d) FFDNet; (e) Ground-truth; (f) DHDN; (g) CMWCNN; (h) CFMNet (Ours).}\label{figkodak24_1}
\end{figure*}

\begin{figure*}[!htbp]	
\tiny
\small
	\scriptsize
	\centering
	\begin{tabular}{cc}
	\tiny
	\scriptsize
	\footnotesize
	\small
		\hspace{-0.4cm}
		\begin{adjustbox}{valign=t}
			\begin{tabular}{cccc}
                \includegraphics[width=0.24\textwidth]{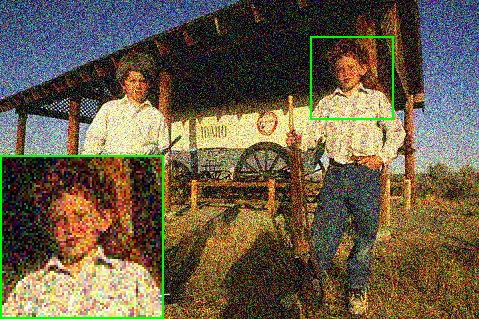}&\hspace{-4.2mm}
				\includegraphics[width=0.24\textwidth]{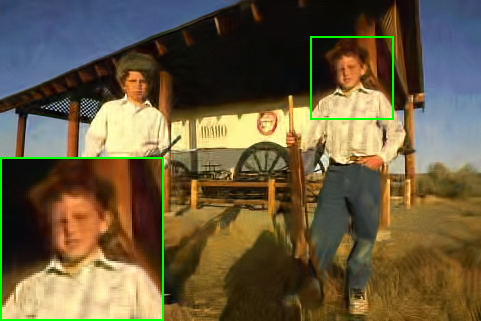}&\hspace{-4.2mm}
				\includegraphics[width=0.24\textwidth]{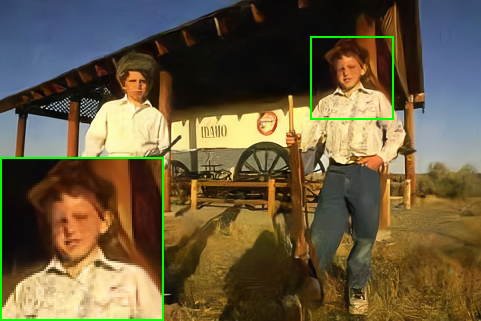}&\hspace{-4.2mm}
				\includegraphics[width=0.24\textwidth]{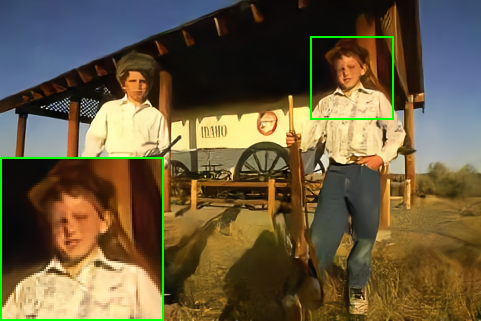}
				\\			
				(a)&\hspace{-4.2mm} 			
				(b)& \hspace{-4.2mm}
				(c)& \hspace{-4.2mm}
				(d)
				\\
				\includegraphics[width=0.24\textwidth]{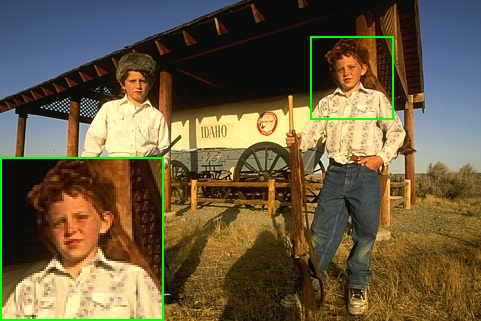}&\hspace{-4.2mm}
				\includegraphics[width=0.24\textwidth]{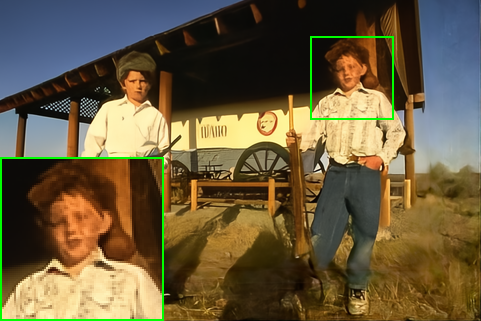}&\hspace{-4.2mm}
				\includegraphics[width=0.24\textwidth]{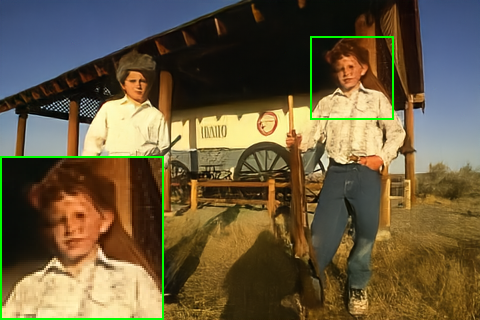}&\hspace{-4.2mm}
				\includegraphics[width=0.24\textwidth]{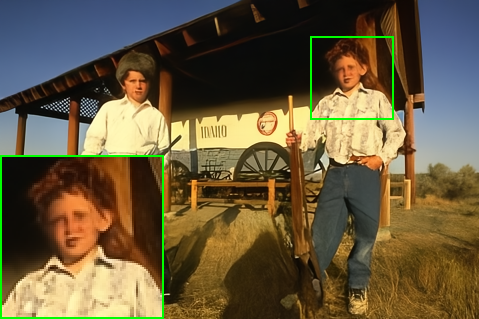}
				\\
                (e)&\hspace{-4.2mm}
				(f)&\hspace{-4.2mm}
				(g)& \hspace{-4.2mm}
				(h)
				\\
			\end{tabular}
		\end{adjustbox}
	\end{tabular}
\caption{Denoising results by different methods on image \emph{test041} from the CBSD68 dataset with $\sigma = 50$. (a) Noisy image; (b) BM3D; (c) DnCNN; (d) FFDNet; (e) Ground-truth; (f) DHDN; (g) CMWCNN; (h) CFMNet (Ours).}\label{figcbsd68_1}
\end{figure*}

\subsection{Experiments on Color Image Denoising}
For color image denoising, we compare our CFMNet with CBM3D~\cite{dabov2007image}, CDnCNN~\cite{zhang2017beyond}, FFDNet~\cite{zhang2018ffdnet}, DHDN~\cite{park2019densely} and CMWCNN~\cite{liu2018multi} on four testing datasets, i.e., CBSD68, Kodak24, McMaster and Urban100.
MemNet~\cite{tai2017memnet}, FOCNet~\cite{jia2019focnet}, N3Net~\cite{Pl2018Neural} and
NLRN~\cite{liu2018non} are not adopted because they did not test their models for color images denoising.
Table~\ref{color_psnr} lists the PSNR and SSIM results for removing AWGN with $\sigma = 15, 25, 50, 75$.
In terms of both PSNR and SSIM, our CFMNet is superior to all the competing methods on the four datasets.
On the McMaster dataset, the PSNR values of our CFMNet are more than 0.3dB higher than those by the second best method, i.e., CMWCNN~\cite{liu2018multi}, for AWGN with any noise levels.
The quantitative results further demonstrate the effectiveness and flexibility of our CFMNet for color image denoising.


Figs.~\ref{figkodak24_1} and~\ref{figcbsd68_1} show the denoising results by different methods on two color images from CBSD68 and Kodak24 with the noise standard deviation $\sigma = 50$.
In Fig.~\ref{figkodak24_1}, our CFMNet can restore more fine details on the eye or the red parrot.
In Fig.~\ref{figcbsd68_1}, more clean details are retained in the face region of the right boy by our CFMNet.
The qualitative results show that our CFMNet is effective in both removing noise and restoring fine-scale image details.


\begin{figure*}[!htbp]
\tiny
\small
	\setlength{\fsfourteen}{-6mm}
	\scriptsize
	\centering
	\begin{tabular}{cc}
	\tiny
	\scriptsize
	\footnotesize
	\small
		\hspace{-0.6cm}
		\begin{adjustbox}{valign=t}
			\begin{tabular}{cccc}
\includegraphics[width=.22\textwidth]{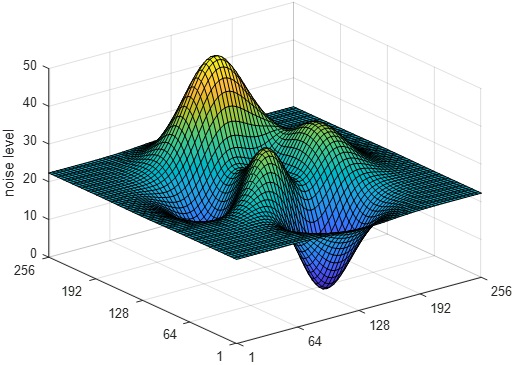}&\hspace{-4.1mm}			
\includegraphics[width=.25\textwidth]{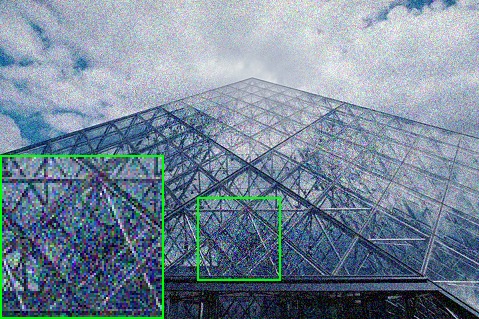}&\hspace{-4.1mm}
\includegraphics[width=.25\textwidth]{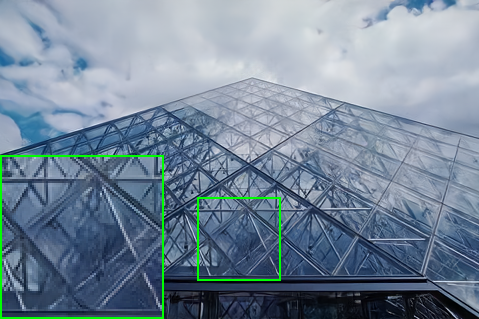}&\hspace{-4.1mm}
\includegraphics[width=.25\textwidth]{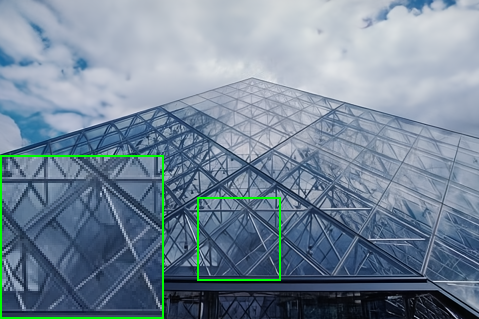}
				\\			
				(a)  &\hspace{-4.1mm}
				(b)  &\hspace{-4.1mm}
                (c) &\hspace{-4.1mm}
				(d)
				\\
			\end{tabular}
		\end{adjustbox}

	\end{tabular}
\caption{Illustration of FFDNet and CFMNet for removing spatially variant AWGN. (a) The noise level map; (b) Noisy image corrupted by spatially variant AWGN; (c) Denoising result by FFDNet (30.15dB/0.9230); (d) Denoising result by our CFMNet (30.91dB/0.9330).}\label{fig_sv1}
\end{figure*}


\subsection{Spatially Variant AWGN Removal.}
\begin{table}[htbp]\footnotesize\arrayrulewidth0.5pt
\centering
\caption{PSNRs/SSIM results of FFDNet and CFMNet for removing spatially variant AWGN}
\label{sv}
\begin{tabular}{|c|c|c|c|}
\hline

Datasets                   & $\sigma$  & FFDNet&CFMNet        \\ \hline
                           & 15    & 37.73/0.9689 & \textbf{38.25/0.9719}          \\
                           & 25    & 34.91/0.9459 & \textbf{35.36/0.9506}         \\
                           & 50    & 31.34/0.8920 & \textbf{31.76/0.9009}          \\
\multirow{-4}{*}{CBSD68}   & 75    & 29.46/0.8455 & \textbf{29.85/0.8583}          \\ \hline
                           & 15    & 38.22/0.9597 & \textbf{38.76/0.9641}          \\
                           & 25    & 35.70/0.9372 & \textbf{36.20/0.9431}          \\
                           & 50    & 32.39/0.8868 & \textbf{32.91/0.8974}         \\
\multirow{-4}{*}{Kodak24}  & 75    & 30.54/0.8442 & \textbf{31.09/0.8590}          \\ \hline
                           & 15    & 37.74/0.9545 & \textbf{38.52/0.9617}          \\
                           & 25    & 35.62/0.9361 & \textbf{36.30/0.9444}          \\
                           & 50    & 32.65/0.8960 & \textbf{33.29/0.9089}          \\
\multirow{-4}{*}{McMaster} & 75    & 30.87/0.8616 & \textbf{31.54/0.8790}          \\ \hline
                           & 15    & 37.24/0.9698 & \textbf{38.01/0.9738}          \\
                           & 25    & 34.84/0.9538 & \textbf{35.59/0.9595}          \\
                           & 50    & 31.65/0.9193 & \textbf{32.53/0.9308}          \\
\multirow{-4}{*}{Urban100} & 75    & 29.76/0.8886 & \textbf{30.77/0.9066}          \\ \hline
\end{tabular}
\end{table}

\begin{table}[htbp]\footnotesize\arrayrulewidth0.5pt
\centering
\caption{Comparison of different CNN denoising models in terms of the number of parameters and running time on images with different size.}
\label{time}
\begin{tabular}{|c|c|c|c|c|}
\hline
\multirow{2}{*}{Methods}&\multirow{2}{*}{\#. Params}&\multicolumn{3}{c|}{\begin{tabular}[c]{@{}c@{}}Running time ($s$)\end{tabular}}\\\cline{3-5}
&  &$256\times256$&$512\times512$&$1024\times1024$\\\hline
DnCNN  & 0.56$M$  & 0.0086& 0.0319 &0.1395\\
FFDNet & 0.49$M$   & 0.0032& 0.0082 &0.0302 \\
N3Net  & 0.71$M$   & 1.0002& 3.9819 &15.9623 \\
DHDN   & 168.19$M$ & 2.0416& 5.9497 &24.6327 \\
MWCNN  & 18.51$M$  & 0.0399& 0.0508&0.1733 \\
CFMNet & 22.56$M$  & 0.0402& 0.1581&0.6261 \\ \hline
\end{tabular}
\end{table}
Among the competing methods, only our CFMNet with FFDNet~\cite{zhang2018ffdnet} are flexible in handling spatially variant AWGN.
First, we adopt the following function to define a spatially variant noise level map,
\begin{align}
{f}(x,y)=3{{(1-x)}^{2}}{{e}^{-{{x}^{2}}-{{(y+1)}^{2}}}} - 10\left(\frac{1}{5}x - {{x}^{3}} - {{y}^{5}}\right) {{e}^{-{{x}^{2}}-{{y}^{2}}}},
\label{eqn_mf}
\end{align}
%
where the variables $x \in [-3, 3]$ and $y \in [-3, 3]$.
Given an image size $H \times W$, the noise level $\tilde{M}_{i,j}$ at position $(i,j)$ is then determined by ${f}(6 \cdot i/H - 3, 6 \cdot j/W - 3)$.
For a noise standard deviation $\sigma = 50$, the final noise level map $\bm{M}$ can be obtained by $\bm{M} = \sigma \cdot (\tilde{\bm{M}} - \min(\tilde{\bm{M}})) / (\max(\tilde{\bm{M}}) - \min(\tilde{\bm{M}}))$ (see Fig.~\ref{fig_sv1}(a)).
To synthesize spatially variant AWGN, we first generate an AWGN noise image $\bf{n}_0$ of zero mean and unit standard deviation, and the spatially variant AWGN can be attained by $\bf{n} = \bf{n}_0 \circ \bm{M}$.
Here, $\circ$ denotes the element-wise product.
Then, the noisy image is obtained by $\bf{x} + \bf{n}$, and Fig.~\ref{fig_sv1}(b) shows an image by adding spatial variant AWGN.

%
%

Table~\ref{sv} lists the average PSNR and SSIM results of FFDNet and CFMNet for handling spatially variant AWGN with noise standard deviation 15, 25, 50 and 75
on four datasets, i.e., CBSD68, Kodak24, McMaster and Urban100.
In terms of both PSNR and SSIM, our CFMNet consistently outperforms FFDNet by a notable margin (i.e., 0.3dB $\sim$ 1dB by PSNR) on all the four datasets.
Fig.~\ref{fig_sv1}(c)(d) shows the denoising results of Fig.~\ref{fig_sv1}(b) by FFDNet and our CFMNet, respectively.
%
Due to their intrinsic flexibility, both FFDNet and CFMNet are able to cope with spatially variant denoising.
Nonetheless, it can be seen that our CFMNet is more promising than FFDNet
in suppressing noise while restoring fine-scale details, further indicating the effectiveness of our CFMNet in removing spatially variant AWGN.

\subsection{Model Size and Running Time}
\begin{table}[htbp]\footnotesize\arrayrulewidth0.5pt
\centering
\caption{Ablation studies on the contributions of (i) feature modulation branch; (ii) short skip connections and (iii) collaboration of convolution activations from noisy image and noise level map.}
\label{ab1}
\begin{tabular}{|c|c|c|c|}
\hline
CFMNet Variants & Running time ($s$) & PSNR/SSIM \\ \hline
CFMNet (w/o CFM)                    &0.0848&29.50/0.8150\\
CFMNet (w/o Res)                    &0.1569&29.56/0.8173\\
CFMNet (w/o ImMod)                 &0.1487&29.54/0.8164\\
CFMNet                            & 0.1621&29.64/0.8195\\ \hline
\end{tabular}
\end{table}
\begin{table}[htbp]\footnotesize\arrayrulewidth0.5pt
\caption{Comparison of CFMNet variants with different feature modulation methods.}
\centering
\label{ab2}
\begin{tabular}{|c|c|c|l|l|}%
\hline
Modulation& Implementation                                   & Time ($s$) & PSNR/SSIM    \\ \hline
CFMNet           &$\bm{f} + \bm{s}$                          &0.1621      &29.64/0.8195\\
CFMNet(scaling) &$\boldsymbol{\gamma} \circ \bm{f}$          &0.1663      &29.66/0.8199\\
CFMNet(affine)   &$\boldsymbol{\gamma} \circ \bm{f} + \bm{s}$&0.1862     &29.66/0.8200\\\hline
\end{tabular}
\end{table}
In addition to quantitative and qualitative evaluation on denoising results, we further compare our CFMNet with several representative deep denoisers by model size and computational efficiency.
We consider six deep CNN denoisers, i.e., DnCNN~\cite{zhang2017beyond}, FFDNet~\cite{zhang2018ffdnet}, N3Net~\cite{Pl2018Neural}, DHDN~\cite{park2019densely}, MWCNN~\cite{liu2018multi} and our CFMNet.
In terms of model size, Table~\ref{time} lists the number of parameters (i.e., \#Params in M) for each method.
{
As for computational efficiency, Table~\ref{time} reports the average running time (in seconds, $s$) on two images from Set12 with size $256 \times 256$ and $512 \times 512$ and one image from Urban100 with size $1024 \times 1024$.
Note that the average running time refers to the GPU running time and is adopted by the average of 10 times calculation.}


Due to the introduction of feature modulation branch, our CFMNet has a larger model size than most competing methods but is much smaller than DHDN~\cite{park2019densely}.
Based on the running time, CFMNet is comparable to DnCNN and MWCNN, and is much more efficient than N3Net and DHDN.

%

\subsection{Ablation Studies}
Using Kodak24 and AWGN with $\sigma = 50$, ablation studies are conducted to assess the effect of major components in CFMNet.
First, to illustrate the contribution of multi-layer CFM modules, we implement a variant of CFMNet, i.e., CFMNet (w/o CFM), by removing the feature modulation branch and taking noisy image and noise level map as the input to the denoising branch.
Moreover, to demonstrate the effect of the residual learning manner in our RS-CFM, a variant of CFMNet, i.e., CFMNet (w/o Res), is implemented by removing the short skip connections adopted in each RS-CFM.
Finally, another variant of CFMNet, i.e., CFMNet (w/o ImMod), is also implemented by only exploiting convolution activations from noise level map to generate shifting map.
Table~\ref{ab1} lists the running time and denoising results of CFMNet and its three variants.
In comparison with CFMNet (w/o CFM), CFMNet can achieve a PSNR gain of 0.14dB, clearly demonstrating the advantage of multi-layer CFM modules over input concatenation.
Comparing CFMNet and CFMNet (w/o Res), the ablation of the short skip connections leads to 0.08dB performance drop showing the effect of the residual learning manner in RS-CFM.
CFMNet also improves CFMNet (w/o ImMod) by 0.1dB in terms of PSNR, indicating that the convolutional activations from  both noise level map and noisy image are beneficial to feature modulation and denoising performance.
In terms of efficiency, CFMNet is inferior to CFMNet (w/o CFM) but is comparable to CFMNet (w/o Res) and CFMNet (w/o ImMod).
\begin{table}[htbp]\footnotesize
\centering
\caption{Comparison of CFMNet variants with different numbers of shifting operations in each CFM module.}
\label{ab3}
\begin{tabular}{|c|c|c|c|c|}
\hline
$\#$. Shifting Operations&  Running Time ($s$) & PSNR/SSIM \\ \hline
1 &0.1142       &29.54/0.8162\\
2 &0.1621       &29.64/0.8195\\
3 &0.2228       &29.64/0.8196\\
4 &0.2788       &29.65/0.8198\\\hline
\end{tabular}
\end{table}

Second, CFMNet uses conditional shifting-based feature modulation for handling AWGN with various noise variance and even spatially variant AWGN.
We also compare other feature modulation methods, such as scaling, and affine transformation including both shifting and scaling.
Table~\ref{ab2} compares CFMNet with its two variants by using scaling (i.e., CFMNet(scaling)) and affine transformation (i.e., CFMNet(affine)) for feature modulation.
CFMNet performs slightly better than CFMNet(scaling) and is comparable to CFMNet(affine).
%
Taking both (i) the tradeoff between denoising result and efficiency and (ii) consistence with input concatenation~\cite{zhang2018ffdnet,mildenhall2018burst} into consideration, shifting-based feature modulation is adopted in our CFMNet.


Finally, each RS-CFM module of CFMNet involves two residual shifting operations.
Obviously, one can adjust the number of residual shifting operations for better balancing denoising performance and efficiency.
Table~\ref{ab3} lists the results of three CFMNet variants with different numbers of residual shifting operations.
As expected, the running time increases along with the increase of number of shifting operations.
In contrast, the PSNR value can be improved by 0.05dB when increasing the number of residual shifting operations from one to two, and then get saturated once it is higher than two.
Thus, we adopt two residual shifting blocks in each RS-CFM in our implementation of CFMNet.

\section{Conclusion}\label{sec:conclusion}
In this paper, we presented a CFMNet by equipping an U-Net backbone with multi-layer residual shifting-based feature modulation (RS-CFM) modules for flexible non-blind image denoising.
CFMNet extended the input concatenation by deploying multiple layers of CFM for better exploiting noise level map to boost denoising performance.
Moreover, each RS-CFM module took the convolutional activations from both noisy image and noise level map as the input to generate shifting map, thereby achieving better tradeoff between noise removal and detail preserving.
Extensive experiments show that our CFMNet performs favorably against the state-of-the-art gray-scale and color image denoising methods, and is effective in handling AWGN with various noise variance and spatially variant AWGN.

\ifCLASSOPTIONcaptionsoff
  \newpage
\fi

\bibliographystyle{IEEEtran}
\bibliography{egbib}

\end{document}